\pdfoutput=1
\documentclass[10pt,a4paper,english,journal]{IEEEtran}
\usepackage[T1]{fontenc}
\usepackage[latin9]{inputenc}
\usepackage{amsmath}
\usepackage{amssymb}
\usepackage{stackrel}
\usepackage{graphicx}

\makeatletter

\pdfpageheight\paperheight
\pdfpagewidth\paperwidth

\providecommand{\tabularnewline}{\\}

\usepackage{xcolor,soul}
\sethlcolor{lightgray}

\@ifundefined{showcaptionsetup}{}{%
 \PassOptionsToPackage{caption=false}{subfig}}
\usepackage{subfig}
\makeatother

\usepackage{babel}
\begin{document}

\title{Semi-Static Radio Frame Configuration for URLLC Deployments in 5G
Macro TDD Networks}

\author{\IEEEauthorblockN{Ali A. Esswie$^{1,2}$,\textit{ }Klaus I. Pedersen\textit{$^{1,2}$,}
and\textit{ }Preben E. Mogensen\textit{$^{1,2}$}\\
$^{1}$Nokia Bell-Labs, Aalborg, Denmark\\
$^{2}$Department of Electronic Systems, Aalborg University, Denmark}}

\maketitle
$\pagenumbering{gobble}$
\begin{abstract}
Dynamic time division duplexing (TDD) is one of the major novelties
of the 5G new radio standard. It notably improves the network resource
utilization with sporadic directional packet arrivals. Although, the
feasibility of the ultra-reliable and low-latency communications (URLLC)
within such deployments is critically challenged, mainly due to the
cross-link interference (CLI). In this work, we propose a semi-static
and computationally-efficient TDD radio frame adaptation algorithm
for 5G macro deployments. Particularly, we first identify the quasi-static
variance of the cross-cell traffic buffering performance, with various
CLI co-existence conditions. Accordingly, a common radio frame pattern
is dynamically estimated based on the filtered multi-cell traffic
statistics. Our system-level simulation results show that the proposed
solution achieves a highly improved URLLC outage performance, i.e.,
offering $\sim40\%$ reduction gain of the achievable URLLC outage
latency compared to perfect static-TDD, and approaching the optimal
interference-free flexible-TDD case; though, with a significantly
lower control overhead size.

\textit{Index Terms}\textemdash{} Dynamic TDD; 5G new radio; URLLC;
Traffic; Cross link interference (CLI).
\end{abstract}

\section{Introduction}

Ultra-reliable and low latency communication (URLLC) is the major
service class of the upcoming fifth generation new radio (5G-NR) standards
{[}1{]}, where it enables a new set of cutting-edge and real-time
applications over wireless mediums, e.g., interactive tactile-internet.
URLLC entails sporadic radio transmissions of a small payload size,
with stringent radio latency and reliability targets of one-way radio
latency of $1$ millisecond with a $99.999\%$ success probability
{[}2{]}. Most of the 5G-NR deployments are envisioned to be with the
time division duplexing (TDD) due to its large spectrum availability
{[}3{]}. Achieving the URLLC targets are particularly challenging
for TDD systems {[}4{]} because of: (a) the non-concurrent downlink
(DL) and uplink (UL) transmission opportunities, and (b) additional
cross-link interference (CLI) among neighboring base-stations (BSs)
adopting opposite transmission directions. Those challenges are particularly
non-trivial for wide-area macro deployments and are the focus of this
paper.

The 5G-NR has defined a flexible slot format design {[}5{]}, where
the TDD adaptation periodicity can be slot-based, i.e., in principal,
per every 14 orthogonal frequency division multiplexing (OFDM) symbols.
Thus, the DL/UL link switching delay is minimized down to less than
a millisecond. However, the CLI still remains a critical capacity
limitation of the flexible TDD deployments. In particular, for a macro
setting, the DL-to-UL CLI, i.e., BS-BS CLI, is most problematic due
to the power imbalance between the DL interfering transmissions and
the UL victim receptions.

In this study, we focus on achieving the URLLC-alike requirements
for macro deployments at frequency range one (FR1), i.e. radio frequency
(RF) operation below 7 GHz. For FR1, neighboring spectrum chunks are
expected to be allocated for different operators. Accordingly, inter-operator
co-existence must be considered, especially to handle the TDD inter-frequency
interference. A study of the 5G-NR TDD RF co-existence was recently
completed by 3GPP, concluding that fully-flexible and uncoordinated
TDD deployments are not possible for FR1 macro deployments due to
the severe BS-BS CLI {[}6{]}, hence, recommending that operators must
adopt fully-aligned TDD radio frame configurations (RFCs) to avoid
the harmful inter-frequency CLI. 

Furthermore, even for single-operator cases, co-channel CLI has been
identified as a severe problem for macro deployments, leading the
use of fully-dynamic TDD to be further challenging. Various methods
to partially handle the co-channel CLI problem have therefore been
proposed in the open literature. Those include CLI cancellation techniques
{[}7-10{]} through inter-cell coordinated user scheduling, joint transceiver
design, power control, and beam-forming. Simpler quasi-dynamic and
opportunistic CLI avoidance schemes are also introduced based on hybrid
RFC design {[}11-13{]}. However, although those techniques offer performance
gain, the CLI problem remains non-negligible, and particularly harmful
for URLLC use cases due to the strict requirements of the achievable
latency and reliability. Needles to say, a simpler, overhead-limited
and CLI-free adaptive RFC selection algorithm is still vital for 5G
macro TDD deployments. 

In this paper, a semi-static and fully-aligned RFC selection algorithm
is proposed for 5G-NR TDD networks. Proposed solution offers CLI-free
TDD transmissions while semi-statically adjusting the RFCs to manage
the tail of the latency-reliability distribution of the experienced
user performance, as the primary performance indicator for URLLC use
cases. The cell-specific traffic load metrics are exchanged across
coordinating cells, and are  filtered to either adapt the upcoming
RFCs to the average or individual cell outage performance. Hence,
the proposed solution adaptively controls the tail distribution of
the cluster capacity and latency, which contributes towards achieving
a decent URLLC outage latency. As the RFC selection is NP-hard problem
for multi-user URLLC deployments, we evaluate the performance of the
proposed solution by means extensive dynamic system-level simulations
to achieve results with high degree of realism. That is, we consider
a dynamic multi-cell, multi-user environment in line with the 5G-NR
specifications, and following the 3GPP simulation modeling guidelines,
e.g. relying on advanced stochastic models for radio propagation,
traffic generation, etc.). Care is taken to achieve trustworthy and
statistical-reliable performance results as a basis for drawing conclusions.

This paper is organized as follows. Section II presents our system
setting. Section III discusses the problem formulation addressed by
this work, while Section IV introduces the proposed scheme. The performance
evaluation appears in Section V. Finally, conclusions are drawn in
Section VI. 

\section{System Model}

We assume a macro 5G-NR TDD network deployment, with $C$ cells, each
with $N_{t}$ antennas. There exits an average number of $K^{\textnormal{dl}}$
and $K^{\textnormal{ul}}$ uniformly-distributed DL and UL active
user-equipment's (UEs) per cell, each equipped with $M_{r}$ antennas.
Herein, we consider the FTP3 traffic model with payload sizes $\textnormal{\ensuremath{\mathit{f}^{dl}}}$
and $\textnormal{\ensuremath{\mathit{f}^{ul}}}$ bits, and Poisson
Point Arrival Process, with mean packet arrivals $\textnormal{\ensuremath{\lambda}}^{\textnormal{dl}}$
and $\textnormal{\ensuremath{\lambda}}^{\textnormal{ul}},$ for DL/UL
links. Hence, the total offered average load per cell is given as:
$\varOmega=\varOmega^{\textnormal{dl}}+\varOmega^{\textnormal{ul}}$,
with $\varOmega^{\textnormal{dl}}=$$K^{\textnormal{dl}}\times\textnormal{\ensuremath{\mathit{f}^{dl}}}\times\textnormal{\ensuremath{\lambda}}^{\textnormal{dl}}$,
and $\varOmega^{\textnormal{ul}}=K^{\textnormal{ul}}\times\textnormal{\ensuremath{\mathit{f}^{ul}}}\times\textnormal{\ensuremath{\lambda}}^{\textnormal{ul}}$
as the average DL and UL offered load sizes, respectively. 

\begin{figure}
\begin{centering}
\includegraphics[scale=0.7]{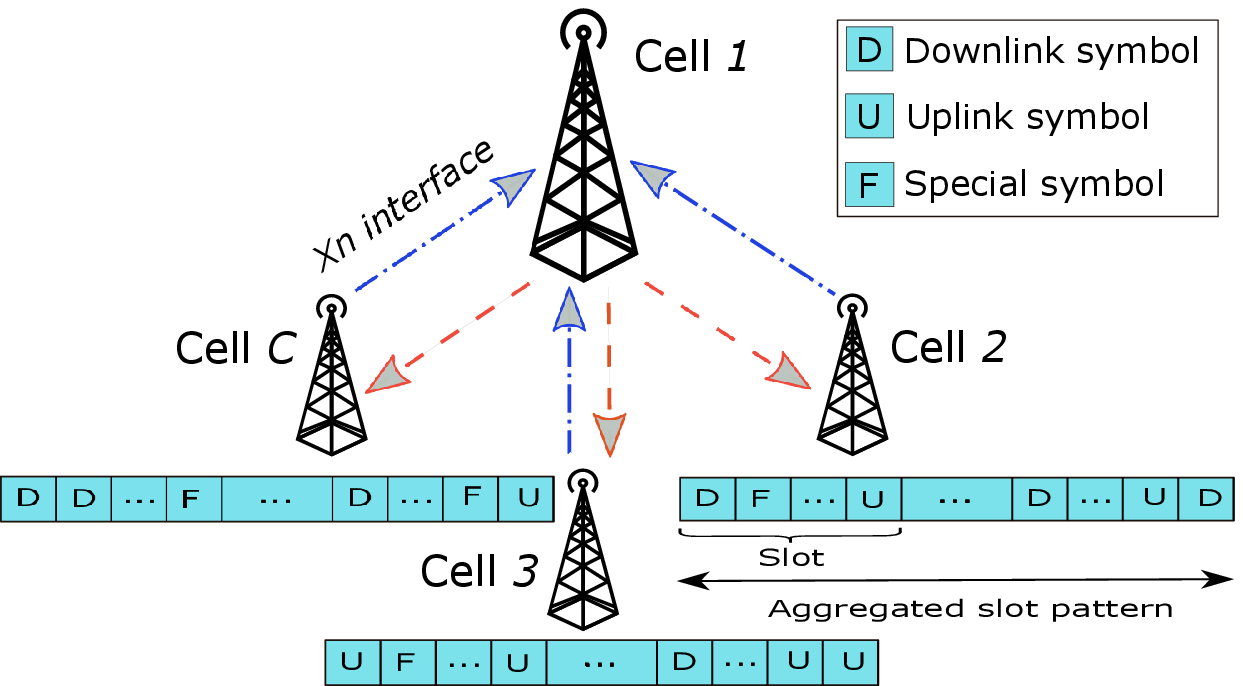}
\par\end{centering}
\centering{}{\small{}Fig. 1. Flexible-TDD network deployment.}{\small \par}
\end{figure}

We follow the latest 3GPP specifications for the 5G-NR TDD system
design. Particularly, the 5G-NR flexible TDD slot format structure
{[}5{]} is considered, as depicted by Fig. 1. A slot format implies
a certain placement of DL {[}D{]}, UL {[}U{]} and flexible {[}F{]}
symbols each 14-OFDM slot duration. In this work, we assume an even
distribution of the DL and UL symbols over the slot in terms of a
4-symbol block size. For instance, a selected slot format with a DL:UL
symbol ratio of 2 : 1 would be: {[}\textbf{DDDDFUUUUDDDDF}{]}. This
configuration allows for sparse DL and UL transmission opportunities
during a slot; though, at the expense of increased guard overhead,
i.e., {[}\textbf{F}{]} symbols. During each slot, UEs are dynamically
multiplexed using the OFDM access (OFDMA), with 30 kHz sub-carrier-spacing
(SCS) and a physical resource block (PRB) of 12 consecutive SCs. The
dynamic user scheduling is performed based on the proportional fair
(PF) criterion, and with a transmit time interval (TTI) duration of
4-OFDM symbols, for rapid URLLC radio transmissions. The achievable
one-way outage URLLC latency is the main performance indicator of
this work. It encompasses the delay from the moment the URLLC packet
becomes available at the packet data convergence protocol (PDCP) layer
until it has been successfully decoded, including the BS and UE processing
delay, hybrid automatic repeat request (HARQ) re-transmission delay,
and scheduling buffering delay, respectively, in line with {[}4{]}.
For UL transmissions, we assume a fast dynamic grant (DG) {[}4{]},
where the UL packets become immediately available for scheduling upon
availability at the UE PDCP layer. That is, the time from transmitting
the UL scheduling request until receiving the DL scheduling grant
is assumed negligible. 

Lets define $\text{\ensuremath{\mathfrak{B}}}_{\textnormal{dl}},$
$\text{\ensuremath{\mathfrak{B}}}_{\textnormal{ul}}$, $\text{\ensuremath{\mathcal{K}}}_{\textnormal{dl}}$
and $\text{\ensuremath{\mathcal{K}}}_{\textnormal{ul}}$ as the sets
of cells and UEs with active DL and UL transmissions, respectively.
Hence, the DL received signal at the $k^{th}$ UE, where $k\text{\ensuremath{\in\text{\ensuremath{\mathcal{K}}}_{\textnormal{dl}}}}$,
$c_{k}\text{\ensuremath{\in\text{\ensuremath{\mathfrak{B}}}_{\textnormal{dl}}}}$,
is given by

\begin{equation}
\boldsymbol{\textnormal{y}}_{k,c_{k}}^{\textnormal{dl}}=\underbrace{\boldsymbol{\textnormal{\textbf{H}}}_{k,c_{k}}^{\textnormal{dl}}\boldsymbol{\textnormal{\textbf{v}}}_{k}s_{k}}_{\text{Useful signal}}+\text{\ensuremath{\mathfrak{T}}}_{k}^{\textnormal{dl}}+\boldsymbol{\textnormal{\textbf{n}}}_{k}^{\textnormal{dl}},
\end{equation}
where $\boldsymbol{\textnormal{\textbf{H}}}_{k,c_{k}}^{\textnormal{dl}}\in\text{\ensuremath{\mathcal{C}}}^{M_{r}\times N_{t}}$
is the DL spatial channel from the cell serving the $k^{th}$ UE,
to the $k^{th}$ UE, $\boldsymbol{\textnormal{\textbf{v}}}_{k}\in\text{\ensuremath{\mathcal{C}}}^{N_{t}\times1}$
, and $s_{k}$ are the single-stream pre-coding vector at the $c_{k}^{th}$
cell, and data symbol of the  $k^{th}$ UE, respectively. $\boldsymbol{\textnormal{\textbf{n}}}_{k}^{\textnormal{dl}}$
is the additive white Gaussian noise, while $\text{\ensuremath{\mathfrak{T}}}_{k}^{\textnormal{dl}}$
denotes the total interference seen at the $k^{th}$ UE receiver end.
Then, $\text{\ensuremath{\mathfrak{T}}}_{k}^{\textnormal{dl}}$ is
expressed by

\begin{equation}
\text{\ensuremath{\mathfrak{T}}}_{k}^{\textnormal{dl}}=\left\{ \begin{array}{c}
\underbrace{\sum_{i\in\text{\ensuremath{\mathcal{K}}}_{\textnormal{dl}}\backslash k}\boldsymbol{\textnormal{\textbf{H}}}_{k,c_{i}}^{\textnormal{dl}}\boldsymbol{\textnormal{\textbf{v}}}_{i}s_{i}}_{\text{DL-to-DL interference}},\,\,\,\,\,\,\textnormal{\,\,\,\,\,\,\,\,\,\,\,\,\,\,\,\,\,\,\,\,\,\,\,\,\,\,\,\,\,\,\,\,\,\,\,\,\,\,Aligned-TDD}\\
\underbrace{\sum_{i\in\text{\ensuremath{\mathcal{K}}}_{\textnormal{dl}}\backslash k}\boldsymbol{\textnormal{\textbf{H}}}_{k,c_{i}}^{\textnormal{dl}}\boldsymbol{\textnormal{\textbf{v}}}_{i}s_{i}}_{\text{DL-to-DL interference}}+\underbrace{\sum_{j\in\text{\ensuremath{\mathcal{K}}}_{\textnormal{ul}}}\boldsymbol{\textnormal{\textbf{G}}}_{k,j}\boldsymbol{\textnormal{\textbf{w}}}_{j}s_{j}}_{\text{UL-to-DL interference}},\,\textnormal{Flexible-TDD}
\end{array}\right.,
\end{equation}
where $\boldsymbol{\textnormal{\textbf{w}}}_{j}\in\text{\ensuremath{\mathcal{C}}}^{M_{r}\times1}$
is the pre-coding vector at the $j^{th}$ UE, and $\boldsymbol{\textnormal{\textbf{G}}}_{k,j}\in\text{\ensuremath{\mathcal{C}}}^{M_{r}\times M_{r}}$
is the the cross-link channel between the $k^{th}$ and $j^{th}$
UEs. Similarly, the received UL signal at the $c_{k}^{th}$ cell,
where $c_{k}\text{\ensuremath{\in\text{\ensuremath{\mathfrak{B}}}_{\textnormal{ul}}}}$
from $k\text{\ensuremath{\in\text{\ensuremath{\mathcal{K}}}_{\textnormal{ul}}}},$
is given as

\begin{equation}
\boldsymbol{\textnormal{y}}_{c_{k},k}^{\textnormal{ul}}=\underbrace{\boldsymbol{\textnormal{\textbf{H}}}_{c_{k},k}^{\textnormal{ul}}\boldsymbol{\textnormal{\textbf{w}}}_{k}s_{k}}_{\text{Useful signal}}+\text{\ensuremath{\mathfrak{T}}}_{c_{k}}^{\textnormal{ul}}+\boldsymbol{\textnormal{\textbf{n}}}_{c_{k}}^{\textnormal{ul}},
\end{equation}
with the total UL interference $\text{\ensuremath{\mathfrak{T}}}_{c_{k}}^{\textnormal{ul}}$
calculated by

\begin{equation}
\text{\ensuremath{\mathfrak{T}}}_{c_{k}}^{\textnormal{ul}}=\left\{ \begin{array}{c}
\underbrace{\sum_{j\in\text{\ensuremath{\mathcal{K}}}_{\textnormal{ul}}\backslash k}\boldsymbol{\textnormal{\textbf{H}}}_{c_{k},j}^{\textnormal{ul}}\boldsymbol{\textnormal{\textbf{w}}}_{j}s_{j}}_{\text{UL-to-UL interference}},\,\,\,\,\,\,\textnormal{\,\,\,\,\,\,\,\,\,\,\,\,\,\,\,\,\,\,\,\,\,\,\,\,\,\,\,\,\,\,\,\,\,\,\,\,\,\,Aligned-TDD}\\
\underbrace{\sum_{j\in\text{\ensuremath{\mathcal{K}}}_{\textnormal{ul}}\backslash k}\boldsymbol{\textnormal{\textbf{H}}}_{c_{k},j}^{\textnormal{ul}}\boldsymbol{\textnormal{\textbf{w}}}_{j}s_{j}}_{\text{UL-to-UL interference}}+\underbrace{\sum_{i\in\text{\ensuremath{\mathcal{K}}}_{\textnormal{dl}}}\boldsymbol{\textnormal{\textbf{Q}}}_{c_{k},c_{i}}\boldsymbol{\textnormal{\textbf{v}}}_{i}s_{i},}_{\text{DL-to-UL interference}}\,\textnormal{Flexible-TDD}
\end{array}\right.,
\end{equation}
where $\boldsymbol{\textnormal{\textbf{Q}}}_{c_{k},c_{i}}\in\text{\ensuremath{\mathcal{C}}}^{N_{t}\times N_{t}}$
is the cross-link channel between the cells serving the $k^{th}$
and $i^{th}$ UEs, $k\text{\ensuremath{\in\text{\ensuremath{\mathcal{K}}}_{\textnormal{ul}}}}$
and $i\in\text{\ensuremath{\mathcal{K}}}_{\textnormal{dl}}$, and
it is measured by orchestrating inter-BS coordinated sounding measurements
{[}10{]}. Accordingly, the achievable post-processing signal-to-interference
(SIR) ratio in the DL direction $\gamma_{k}^{\textnormal{dl}}$ and
UL direction $\gamma_{c_{k}}^{\textnormal{ul}}$ are given by

{\small{}
\begin{equation}
\gamma_{k}^{\textnormal{dl}}=\frac{\left\Vert \left(\boldsymbol{\textnormal{\textbf{u}}}_{k}^{\textnormal{dl}}\right)^{\textnormal{H}}\boldsymbol{\textnormal{\textbf{H}}}_{k,c_{k}}^{\textnormal{dl}}\boldsymbol{\textnormal{\textbf{v}}}_{k}\right\Vert ^{2}}{\underset{i\in\text{\ensuremath{\mathcal{K}}}_{\textnormal{dl}}\backslash k}{\sum}\left\Vert \left(\boldsymbol{\textnormal{\textbf{u}}}_{k}^{\textnormal{dl}}\right)^{\textnormal{H}}\boldsymbol{\textnormal{\textbf{H}}}_{k,c_{i}}^{\textnormal{dl}}\boldsymbol{\textnormal{\textbf{v}}}_{i}\right\Vert ^{2}+\underset{j\in\text{\ensuremath{\mathcal{K}}}_{\textnormal{ul}}}{\sum}\left\Vert \left(\boldsymbol{\textnormal{\textbf{u}}}_{k}^{\textnormal{dl}}\right)^{\textnormal{H}}\boldsymbol{\textnormal{\textbf{G}}}_{k,j}\boldsymbol{\textnormal{\textbf{w}}}_{j}\right\Vert ^{2}},
\end{equation}
}{\small \par}

{\small{}
\begin{equation}
\gamma_{c_{k}}^{\textnormal{ul}}=\frac{\left\Vert \left(\boldsymbol{\textnormal{\textbf{u}}}_{k}^{\textnormal{ul}}\right)^{\textnormal{H}}\boldsymbol{\textnormal{\textbf{H}}}_{c_{k},k}^{\textnormal{ul}}\boldsymbol{\textnormal{\textbf{w}}}_{k}\right\Vert ^{2}}{\underset{j\in\text{\ensuremath{\mathcal{K}}}_{\textnormal{ul}}\backslash k}{\sum}\left\Vert \left(\boldsymbol{\textnormal{\textbf{u}}}_{k}^{\textnormal{ul}}\right)^{\textnormal{H}}\boldsymbol{\textnormal{\textbf{H}}}_{c_{k},j}^{\textnormal{ul}}\boldsymbol{\textnormal{\textbf{w}}}_{j}\right\Vert ^{2}+\underset{i\in\text{\ensuremath{\mathcal{K}}}_{\textnormal{dl}}}{\sum}\left\Vert \left(\boldsymbol{\textnormal{\textbf{u}}}_{k}^{\textnormal{ul}}\right)^{\textnormal{H}}\boldsymbol{\textnormal{\textbf{Q}}}_{c_{k},c_{i}}\boldsymbol{\textnormal{\textbf{v}}}_{i}\right\Vert ^{2}},
\end{equation}
}with $\left\Vert \bullet\right\Vert ^{2}$ as the second-norm, and
$\boldsymbol{\textnormal{\textbf{u}}}_{k}^{\kappa}\in\text{\ensuremath{\mathcal{C}}}^{N_{t}/M_{r}\times1}$,
$\text{\ensuremath{\mathcal{X}}}^{\kappa},\kappa\text{\ensuremath{\in}}\{\textnormal{ul},\textnormal{dl}\}$,
is the linear minimum mean square error interference rejection combining
(LMMSE-IRC) receiver {[}14{]}, and $\left(\bullet\right)^{\textnormal{H}}$
denotes the Hermitian operation. 

\section{Problem Formulation }

The URLLC outage performance is dominated by the achievable radio
latency at the lower $10^{-5}$ outage probability. This implies a
stringent latency bound with a rare violation occurrence. Thus, in
TDD deployments, due to the non-concurrent DL and UL transmission,
the URLLC latency and reliability targets become highly susceptible
to the number and placement of the DL $d_{c}$ and UL $u_{c}$ symbols
during an RFC. Accordingly, our objective is to optimize the RFC selection
in order to minimize the URLLC outage radio latency as

\begin{equation}
\begin{array}{c}
\left(\frac{\textnormal{\ensuremath{d_{c}}}}{u_{c}}\right)^{*}\triangleq\left\{ \frac{\textnormal{\ensuremath{d^{i}}}}{u^{i}}\,\,:\,\,\frac{\textnormal{\ensuremath{d^{i}}}}{u^{i}}\in\text{\ensuremath{\mathfrak{T}}}\right\} \\
\\
\textnormal{s.t:}\underset{k}{\,\,\,\arg\min}\left(\varphi_{c,k}\right),\,\text{\ensuremath{\forall}}k\in\text{\ensuremath{\mathcal{K}_{\textnormal{ul}/\textnormal{dl}}}},
\end{array}
\end{equation}
where $\text{\ensuremath{\mathfrak{T}}}$ is the set of all pre-defined
possible RFC structures, $\varphi_{c,k}$ is the one-way URLLC radio
latency {[}4{]}. Accordingly, there is no feasible optimal solution
of the DL $d_{c}^{\textnormal{opt.}}$ and UL $u_{c}^{\textnormal{opt.}}$
symbol structure to satisfy the UE-specific latency and reliability
requirements. For instance, in multi-UE URLLC deployments, and due
to the time-variant sporadic traffic arrivals, multiple UEs may request
simultaneous opposite link directions. Thus, BSs instead adapt the
RFC structure, on a best effort basis, to offer faster transmissions
of the UEs with the worst latency performance while buffering other
UEs. Adding the severe BS-BS CLI on top, victim UL packets most likely
inflict several HARQ re-transmissions before a successful decoding,
violating the UE-specific latency budget as well as dictating the
RFC adaptation by pending packets rather than the new packet arrivals.
Thus, to tackle this issue, we propose a semi-static coordinated RFC
selection algorithm, which offers fully CLI-free transmissions while
adapting the RFC selection to the varying URLLC latency statistics. 

\section{Proposed Coordination Scheme}

We propose a computationally-efficient RFC selection algorithm to
offer a decent URLLC outage latency performance. First, cells estimate
their average directional traffic size on a pre-defined periodicity.
Then, a relative load metric is shared among the coordinating cells
over the back-haul Xn-interface, i.e., multiple bits of feedback.
Subsequently, a filtering window is applied on the reported traffic
data-set, either to match the average traffic volume per cluster,
i.e., equal-priority windowing, or biasing the RFC adaptation towards
individual cells, e.g., typically those with the worst traffic buffering
performance. Then, a common RFC is estimated to match the filtered
traffic volume, and accordingly, cells within the cluster adopt the
same RFC until the next RFC update instant. 

\subsection{Cell-specific directional traffic tracking}

Cells seek to select the RFCs which minimize the achievable average
URLLC outage latency, according to (7). In multi-user URLLC networks,
there may exist several active UEs with simultaneous UL and DL transmission
requests, respectively, and hence, cross-directional target latency
conflict is exhibited. 

For the considered URLLC use cases, the incoming traffic is only allowed
to be buffered for a short time-duration before being transmitted.
Otherwise, the URLLC latency constraint is violated. We can therefore
observe the strong correlation between the amount of buffered data,
i.e., queuing of data payloads, and the experienced latency. Selecting
the RFC that offers the best URLLC outage performance is therefore
translated to selecting the RFC which minimizes the DL and UL buffering.
As the offered traffic increases, buffering is obviously unavoidable,
thus, the best feasible solution from the RFC selection point of view
is to ensure that the traffic buffering of the two link directions
is balanced. 

Accordingly, at the $\varrho^{th}$ slot of the radio frame, $\varrho=1,2,\ldots,\rho$,
with $\rho$ as the number of slots per radio frame, the $c^{th}$
cell calculates the aggregated DL $Z_{c}^{\textnormal{dl}}\left(\varrho\right)$
and UL $Z_{c}^{\textnormal{ul}}\left(\varrho\right)$ buffered traffic
size, respectively. Specifically, the UL traffic volume is identified
at the cell side from the UE scheduling requests, and the associated
buffer status reports. Thus, the normalized traffic ratio $\mu_{c}\left(\varrho\right)$
is defined as 

{\small{}
\begin{equation}
\mu_{c}\left(\varrho\right)=\frac{Z_{c}^{\textnormal{dl}}\left(\varrho\right)}{Z_{c}^{\textnormal{dl}}\left(\varrho\right)+Z_{c}^{\textnormal{ul}}\left(\varrho\right)}.
\end{equation}
}{\small \par}

Then, the instantaneous traffic ratios $\mu_{c}\left(\varrho\right)$
are linearly averaged across each frame duration as expressed by

\begin{equation}
\overline{\mu}_{c}=\frac{1}{\rho}\,\stackrel[\varrho=1]{\rho}{\sum}\mu_{c}\left(\varrho\right),
\end{equation}
where $\overline{\mu}_{c}$ implies the averaged traffic ratio of
the $c^{th}$ cell. In case there are neither DL and UL new packet
arrivals or buffered traffic, BSs fall-back to a default RFC structure
until the next update instant. The larger the $\overline{\mu}_{c}$,
e.g., $\sim1$, the larger the buffered DL traffic compared to the
corresponding UL traffic volume. For instance, with $\overline{\mu}_{c}=0.9$,
the DL traffic volume is $\textnormal{9x}$ the respective UL volume.
Finally, per every RFC adaptation period, neighboring cells exchange
the measured $\overline{\mu}_{c}$ over the Xn-interface. 

\subsection{Traffic filtering and common RFC selection }

As the URLLC outage performance is mainly dominated by the cells of
the worst latency and reliability performance, we apply a window filtering
to dynamically control the URLLC latency tail distribution, i.e.,
outage latency. Thus, based on the exchange of $\overline{\mu}_{c}$,
the most problematic cells are first identified as those inflicting
the largest or smallest $\overline{\mu}_{c}$, i.e., having too large
DL or UL buffered traffic volume, which implies the RFC adaptation
is not properly configured for those victim cells. In this regard,
cells calculate the absolute linear distance of the reported $\overline{\mu}_{c}$
data-set towards its mean value $d$, i.e., $d=0.5$, and then, sort
them in an descending order in terms of their respective absolute
linear distance, as given by

\begin{equation}
\boldsymbol{\Psi}=\underset{x^{\left(a\right)},\,y^{\left(b\right)};\,a>b}{\textnormal{\textbf{Sort}}}\left[\overline{\mu}_{1}^{\left(\left|\overline{\mu}_{1}-d\right|\right)},\,\overline{\mu}_{2}^{\left(\left|\overline{\mu}_{2}-d\right|\right)},\,\ldots,\,\overline{\mu}_{C}^{\left(\left|\overline{\mu}_{C}-d\right|\right)}\right].
\end{equation}

The ordered traffic data-set $\Psi=\left[\psi_{1},\,\psi_{2},\,\ldots,\,\psi_{C}\right]$
is then filtered using a spatial window. In this work, we consider
the Kaiser window $w\left[l\right]$, due to its flexible response
tunability, and is given in the discrete domain as

\begin{equation}
w\left[l\right]=\frac{I_{0}\left[\beta\sqrt{1-\left(\frac{2l}{L}-1\right)^{2}}\right]}{I_{0}\left[\beta\right]},\,0\leq l\leq L
\end{equation}
with $I_{0}$ as the zeroth-order modified Bessel function of the
first kind, $\beta$ is the window shaping factor, and $L+1$ denotes
the window length. As depicted by Fig. 2, the mirrored Kaiser window
amplitude is shown for various normalized shaping factors $\hat{\beta}=\frac{\beta}{\beta_{\textnormal{max}}}$,
with $\beta_{\textnormal{max}}=100$ and $L=100$. The larger the
$\hat{\beta}$ factor, the more selective the Kaiser window. For instance,
with $\hat{\beta}=0$, the mirrored Kaiser window approaches a conventional
band pass filter, i.e., all cells are equally prioritized; although,
with a larger $\hat{\beta}=0.9,$ the window becomes highly selective
over a subset of the sample space, i.e., certain cells are highly
prioritized. 

\begin{figure}
\begin{centering}
\includegraphics[scale=0.6]{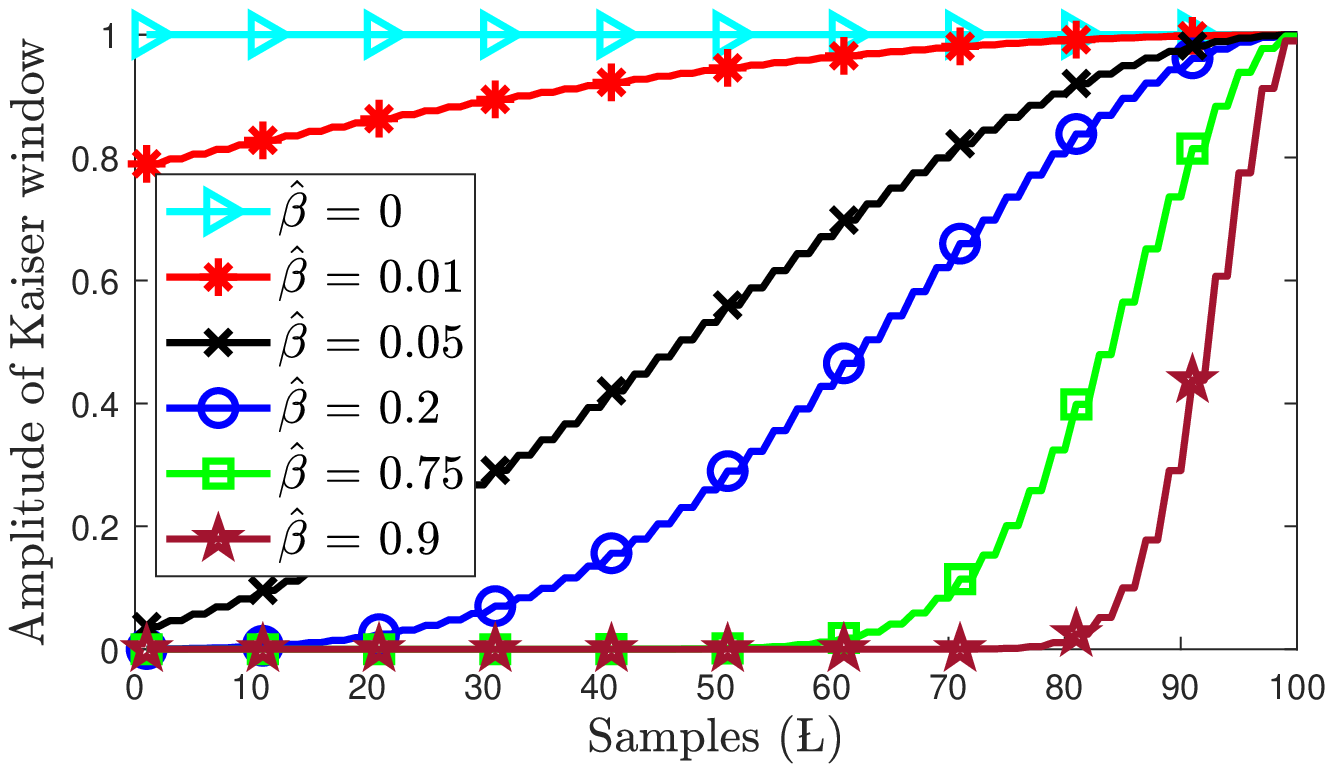}
\par\end{centering}
\centering{}{\small{}Fig. 2. Mirrored Kaiser window with $\hat{\beta}$.}{\small \par}
\end{figure}

Accordingly, the Kaiser window coefficients are applied in a descending
order on the sorted data-set $\Psi$, as 

\begin{equation}
\varTheta=\frac{\psi_{1}w\left[0\right]+\psi_{2}w\left[1\right]+\cdots+\psi_{C}w\left[L\right]}{w\left[0\right]+w\left[1\right]+\cdots+w\left[L\right]},\,L=C-1,
\end{equation}
where $\varTheta$ is the filtered traffic ratio per the entire cluster,
with $w\left[0\right]>w\left[1\right]>\cdots>w\left[L\right]$. Based
on the calculated $\varTheta,$ a common RFC is selected and adopted
by all cells within the cluster until the next RFC update instant.
For example, with an estimated $\varTheta=0.2$, a common RFC of $d_{c}/u_{c}\simeq\frac{1}{4}$
is adopted across all cells, with the DL/UL symbol placement configured
according to the strategy presented in Section II. 

\subsection{Comparison to the state-of-the-art TDD studies}

We compare the performance of the proposed solution against the state-of-the-art
TDD solutions in the recent literature as follows:

\textbf{Static-TDD (sTDD)}: a pre-defined RFC is globally configured
for all cells across the entire network, where it matches the average
network traffic demand. Herein, we define $\alpha$ as the normalized
RFC mismatch error, where $\alpha=0$ implies the global RFC is selected
to perfectly match the average network traffic statistics and $\alpha=0.35$
denotes $35\%$ symbol mismatch of the configured RFC against the
actual average offered traffic load. sTDD deployments offer CLI-free
conditions; though, with a limited cross-cell traffic adaptation flexibility.

\textbf{Dynamic TDD (dTDD)}: a fully flexible TDD operation is assumed,
where at each RFC update period, each cell independently adopts the
RFC which best meets its individual traffic demand. We consider two
scenarios of the dTDD deployments as: (a) a dTDD setting with an optimal
CLI cancellation (dTDD-CLI-free) {[}9{]}, where the BS-BS and UE-UE
are perfectly suppressed using full packet exchange over both the
back-haul and radio interfaces, and (b) a dTDD deployment with CLI
coexistence (dTDD-CLI). 
\begin{table}
\caption{{\small{}Simulation parameters.}}
\centering{}%
\begin{tabular}{c|c}
\hline 
Parameter & Value\tabularnewline
\hline 
Environment & 3GPP-UMA, one cluster, 21 cells\tabularnewline
\hline 
UL/DL channel bandwidth & 10 MHz, SCS = 30 KHz, TDD\tabularnewline
\hline 
Carrier frequency & 3.5 GHz\tabularnewline
\hline 
TDD mode & Synchronized \tabularnewline
\hline 
Antenna setup & $N_{t}=8$ , $M_{r}=2$ \tabularnewline
\hline 
Average user load per cell & $K^{\textnormal{dl}}=K^{\textnormal{ul}}=$ 10 \tabularnewline
\hline 
TTI duration & 4-OFDM symbols\tabularnewline
\hline 
Traffic model & $\begin{array}{c}
\textnormal{FTP3, \textnormal{\ensuremath{\mathit{f}^{dl}}} = \textnormal{\ensuremath{\mathit{f}^{ul}}} = 400 bits}\\
\textnormal{\ensuremath{\textnormal{\ensuremath{\lambda}}^{\textnormal{dl}}} =125, and 375 pkts/sec}\\
\textnormal{\ensuremath{\textnormal{\ensuremath{\lambda}}^{\textnormal{ul}}} =125, and 375 pkts/sec}
\end{array}$\tabularnewline
\hline 
Offered load ratio  & $\textnormal{\textnormal{DL:UL = 1:1}}$\tabularnewline
\hline 
Processing time & $\begin{array}{c}
\textnormal{Preperation delay: 3-OFDM symbols }\\
\textnormal{PDSCH decoding: 4.5-OFDM symbols}\\
\textnormal{PUSCH decoding : 5.5-OFDM symbols}
\end{array}$\tabularnewline
\hline 
RFC update periodicity & 10 ms (radio frame)\tabularnewline
\hline 
UL/DL receiver & LMMSE-IRC\tabularnewline
\hline 
Link adaptation & Adaptive modulation and coding\tabularnewline
\hline 
HARQ configuration & asynchronous with Chase Combining\tabularnewline
\hline 
\end{tabular}
\end{table}

\section{Performance Evaluation}

We assess the performance of the proposed solution using extensive
system-level simulations, with a high degree of realism. The major
simulation settings are listed in Table I, where the main assumptions
of the 3GPP release-15 for TDD deployments are adopted. During every
RFC update periodicity, cells estimate their buffered traffic ratio,
according to (8), and hence, share it among the cluster in order to
estimate a common RFC. Thus, during each TTI, DL and UL UEs are dynamically
multiplexed using OFDMA based on the PF metric. The signal-to-interference-noise-ratio
(SINR) points of the individual SCs are calculated by the LMMSE-IRC
receiver, and combined into an effective SINR level using the exponential
SNR mapping {[}15{]}. Finally, we adopt a dynamic link adaptation,
i.e., adaptive modulation and coding selection, and asynchronous HARQ
Chase combining, where the HARQ re-transmissions are dynamically scheduled,
and are always prioritized over new transmissions. 

Fig. 3 depicts the complementary cumulative distributive function
(CCDF) of the DL/UL combined URLLC one-way radio latency in ms, of
the proposed scheme, sTDD, and dTDD, respectively, for{\small{} $\varOmega=1$
Mbps.} Looking at the achievable latency at the $10^{-5}$ probability
level, i.e., URLLC outage latency, the proposed solution clearly offers
a decent URLLC outage performance, approaching the optimal dTDD-CLI-free.
It achieves $\sim20\%$ and $\sim40\%$ reduction of the URLLC outage
latency, compared to the sTDD with a perfect RFC match, i.e., $\alpha=0$
and non-perfect RFC match, i.e., $\alpha=0.35$, respectively. However,
the proposed solution inflicts $\sim10\%$ URLLC outage latency degradation
against the ideal dTDD-CLI-free; although, this comes with a significantly
lower control signaling overhead size. In that sense, the proposed
solution relaxes one of the most challenging requirements of the conventional
sTDD schemes, as it does not require a pre-configured RFC, while still
preserving fully CLI-free transmissions with a semi-static traffic
adaptation. 

Furthermore, the dTDD-CLI-free achieves the best URLLC outage latency,
i.e., $\sim1.78$ ms, due to the fully flexible, i.e., cell-wise,
RFC adaptation to the individual sporadic traffic arrivals. Though,
it comes with the assumption of optimal CLI-free conditions. The dTDD-CLI
scheme exhibits an outage latency saturation, since the URLLC radio
performance becomes dictated by the aggressive BS-BS CLI, instead
of the RFC adaptation, leading to the consumption of the maximum HARQ
attempts before UL packets are either dropped or successfully received
after the Chase combining HARQ process. 

\begin{figure}
\begin{centering}
\includegraphics[scale=0.5]{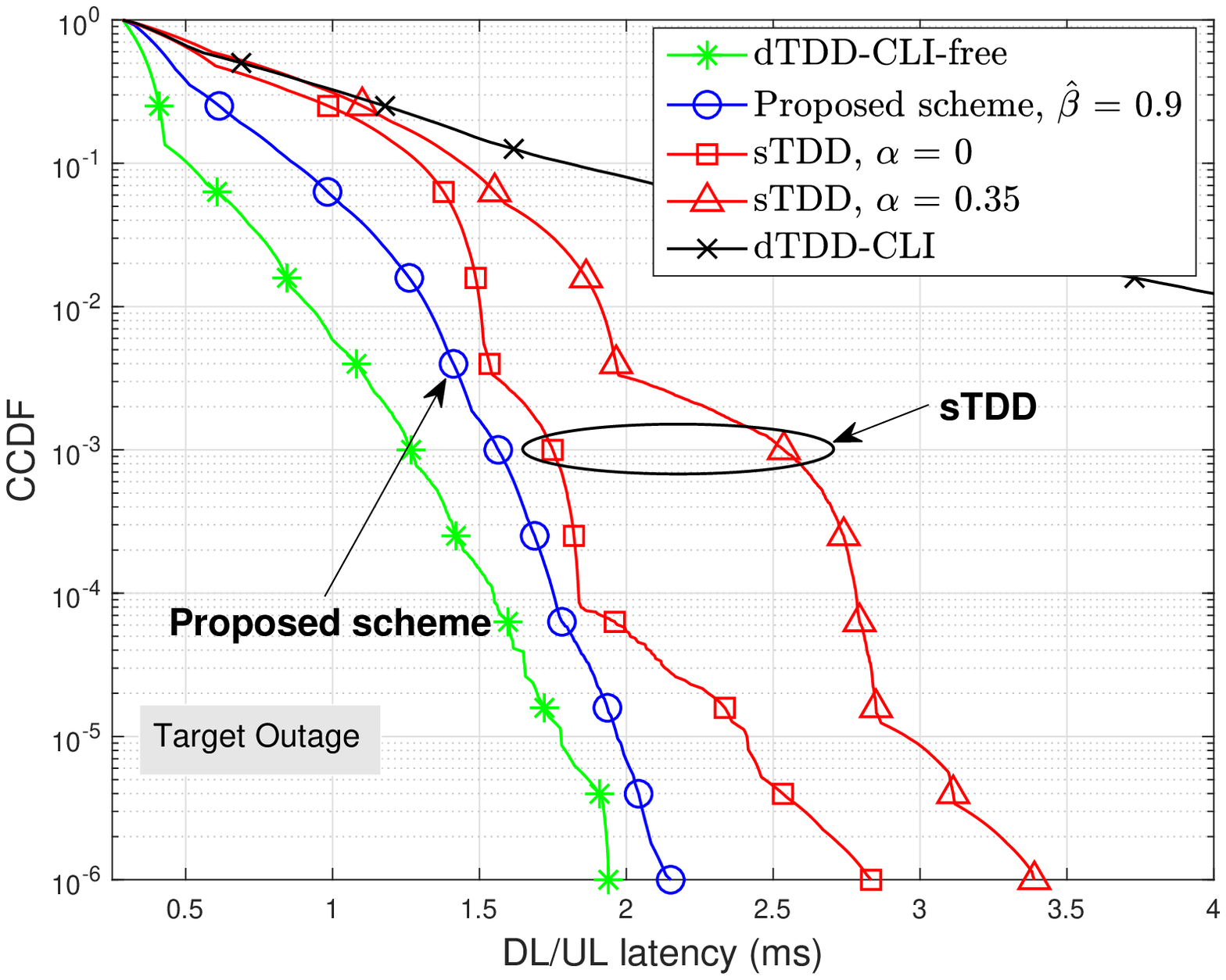}
\par\end{centering}
\centering{}{\small{}Fig. 3. Achievable URLLC outage latency of proposed
scheme, sTDD, and dTDD, with $\varOmega=1$ Mbps.}{\small \par}
\end{figure}
 
\begin{figure}
\begin{centering}
\includegraphics[scale=0.5]{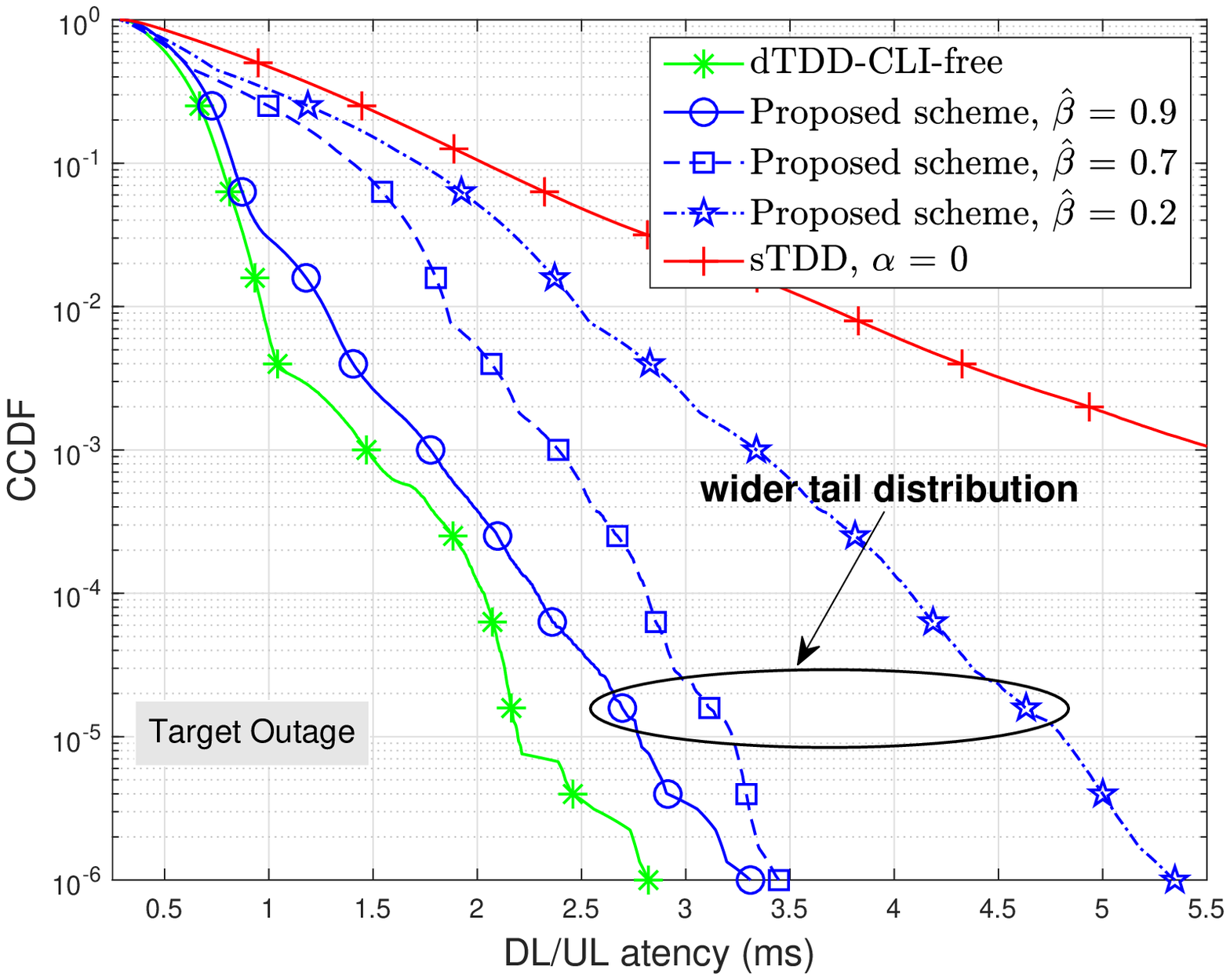}
\par\end{centering}
\centering{}{\small{}Fig. 4. Achievable URLLC outage latency of proposed
scheme, with $\hat{\beta}$ and $\varOmega=3$ Mbps.}{\small \par}
\end{figure}

Fig. 4 shows the CCDF of the URLLC latency of the proposed solution,
with different $\hat{\beta}$ settings, and $\varOmega=3$ Mbps. As
can be noticed, with $\hat{\beta}=0.9$, the tail of the URLLC latency
distribution becomes more narrower, since the cells with the worst
traffic buffering performance are highly prioritized in selecting
the upcoming RFCs. With a smaller $\hat{\beta}$ factor, the reported
traffic statistics of the coordinating cells are equally prioritized,
leading to a wider latency tail distribution. That is $\sim+54\%$
increase in the URLLC outage latency with $\hat{\beta}=0.2$, compared
to the case with $\hat{\beta}=0.9$. This consolidates the fact that
the URLLC outage latency is dictated by the cells of the worst buffering
imbalance. Hence, those should be given a higher priority when deciding
the upcoming RFCs, in order to rapidly recover their respective outage
targets. 

Looking at the URLLC outage performance with different $\varOmega^{\textnormal{dl}}/\varOmega^{\textnormal{ul}}$
ratios, Fig. 5 depicts the CCDF of the URLLC radio latency, under
the proposed and the sTDD schemes, respectively, where the latter
is configured with $d_{c}/u_{c}=1$, for $\varOmega^{\textnormal{dl}}/\varOmega^{\textnormal{ul}}=3:1$
and $1:3$. The proposed solution offers a sufficient frame adaptation
against the variable offered traffic ratio $\varOmega^{\textnormal{dl}}/\varOmega^{\textnormal{ul}}$,
resulting in a decent URLLC outage latency, i.e., $3.8$ ms for $\varOmega^{\textnormal{dl}}/\varOmega^{\textnormal{ul}}=3:1$
and $2.9$ ms for $\varOmega^{\textnormal{dl}}/\varOmega^{\textnormal{ul}}=1:3$,
respectively. The sTDD scheme clearly exhibits a significant outage
latency increase due to the mismatch between the predefined $d_{c}/u_{c}=1$
and the offered traffic ratio $\varOmega^{\textnormal{dl}}/\varOmega^{\textnormal{ul}}$,
e.g., proposed solution offers $\sim82\%$ outage latency reduction
compared to the sTDD scheme. Accordingly, the proposed solution eliminates
the rigid requirement of pre-configuring a global RFC while offering
a semi-static RFC adaptation to the varying offered traffic. 

\begin{figure}
\begin{centering}
\subfloat[DL-heavy ($\varOmega^{\textnormal{dl}}:\varOmega^{\textnormal{ul}}=3:1$)]{\begin{centering}
\includegraphics[scale=0.5]{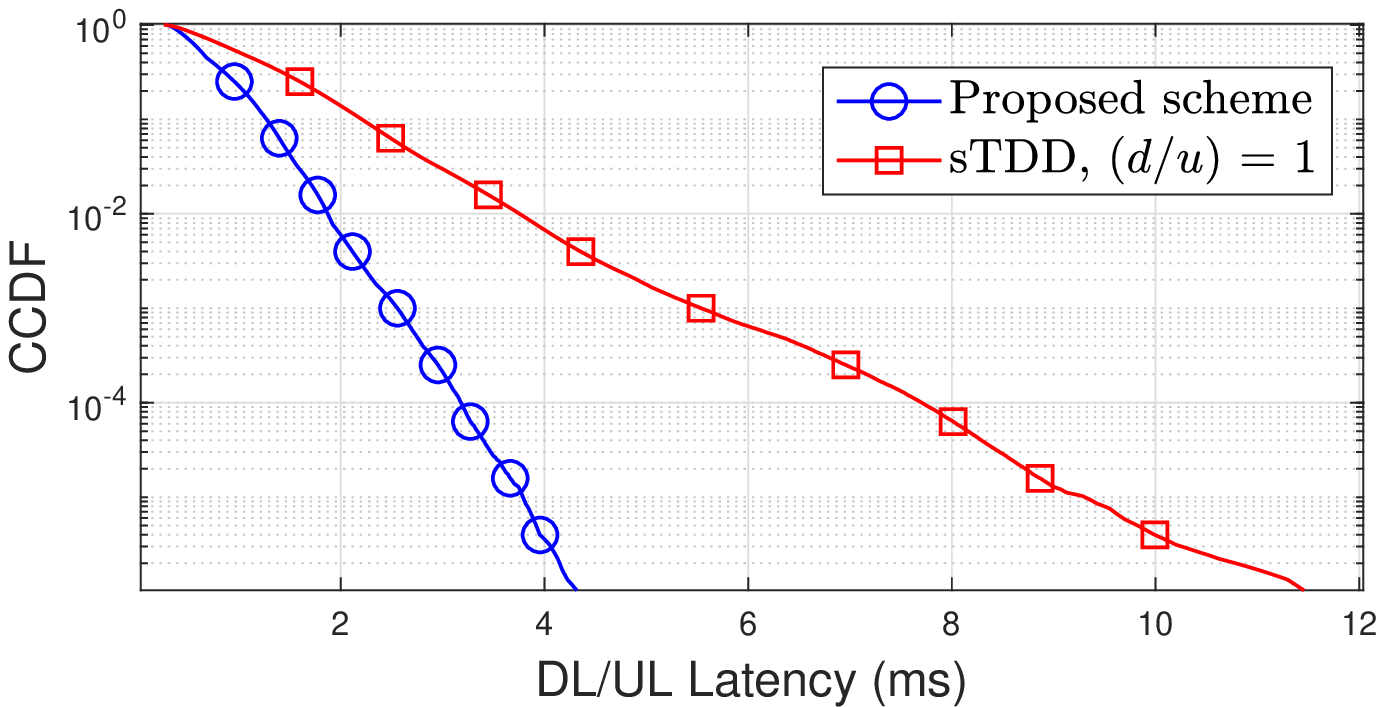}
\par\end{centering}
}
\par\end{centering}
\begin{centering}
\subfloat[UL-heavy ($\varOmega^{\textnormal{dl}}:\varOmega^{\textnormal{ul}}=1:3$)]{\begin{centering}
\includegraphics[scale=0.5]{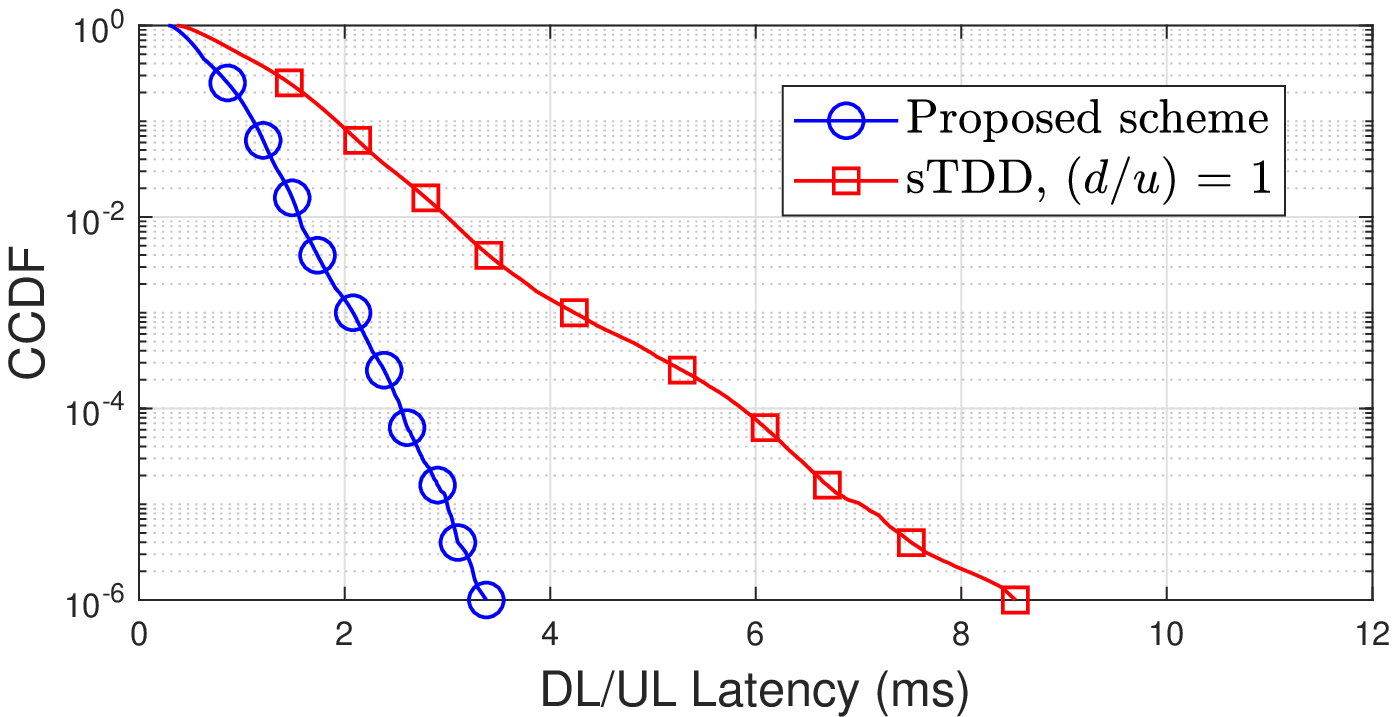}
\par\end{centering}
}
\par\end{centering}
\centering{}{\small{}Fig. 5. Comparison of the URLLC latency performance
with various DL and UL traffic ratios, }$\varOmega=3$ Mbps{\small{}.}{\small \par}
\end{figure}

Finally, Fig. 6 depicts a comparison of the achievable combined DL/UL
outage latency with various offered load levels, in reference to the
dTDD-CLI-free scheme. At the very low offered region {\small{}$\varOmega=0.25$
Mbps}, with an average of a single active UE per cell, the traffic
demand becomes highly variant among neighboring cells. Accordingly,
the URLLC outage performance is dominated by how fast the cells adapt
their individual RFCs to the sporadic traffic arrivals. Thus, both
the proposed solution and sTDD schemes inflict a considerable outage
latency degradation, compared to the optimal dTDD-CLI-free, i.e.,
$+29\%$ and $+63\%$ latency increase. However, the proposed scheme
outperforms the corresponding sTDD by $35\%$ outage latency reduction
gain. This is mainly attributed to the semi-static cross-cell RFC
adaptation of the proposed solution. Thus, unlike sTDD, cells with
accumulating traffic size are given higher priority in selecting the
RFCs. Over the high load region {\small{}$\varOmega=5$ Mbps}, similar
conclusions are observed; although, with less outage latency relative
degradation compared to the optimal dTDD-CLI-free, since in this case,
the URLLC outage performance is mainly dictated by the scheduling
 queuing delay rather than the flexibility of the RFC adaptation operation. 

\section{Concluding Remarks }

A semi-static radio frame configuration (RFC) selection algorithm
has been proposed for 5G TDD macro deployments. Proposed solution
incorporates a simple inter-cell signaling exchange procedure of the
relative traffic statistics, in order to estimate a common RFC of
each cluster, which matches the time-variant and cell-specific traffic
demand. Compared to the state-of-the-art TDD literature, the proposed
solution demonstrates an attractive trade-off between the achievable
URLLC outage performance and the signaling overhead size. It achieves
$\sim40\%$ reduction of the URLLC outage latency, compared to the
ideal static-TDD deployment, while approaching the optimal dynamic-TDD
bound; though, with a significantly lower signaling overhead size. 

The main insights brought by this paper are as follows: (a) within
macro 5G new radio deployments, the cross-cell traffic statistics
are of a low time-variance due to the sufficiently large number of
active connected users, (b) accordingly, relaxing the requirements
of fast traffic adaptation for the sake of controlling the critical
cross-link interference (CLI) is of a more significance, and (c) the
proposed solution offers a flexible and semi-static RFC adaptation
to the sporadic cross-cell traffic demand, with fully CLI-free conditions
and limited signaling overhead. 

\section{Acknowledgments}

This work is partly funded by the Innovation Fund Denmark \textendash{}
File: 7038-00009B. The authors would like to acknowledge the contributions
of their colleagues in the project, although the views expressed in
this contribution are those of the authors and do not necessarily
represent the project. 
\begin{figure}
\begin{centering}
\includegraphics[scale=0.58]{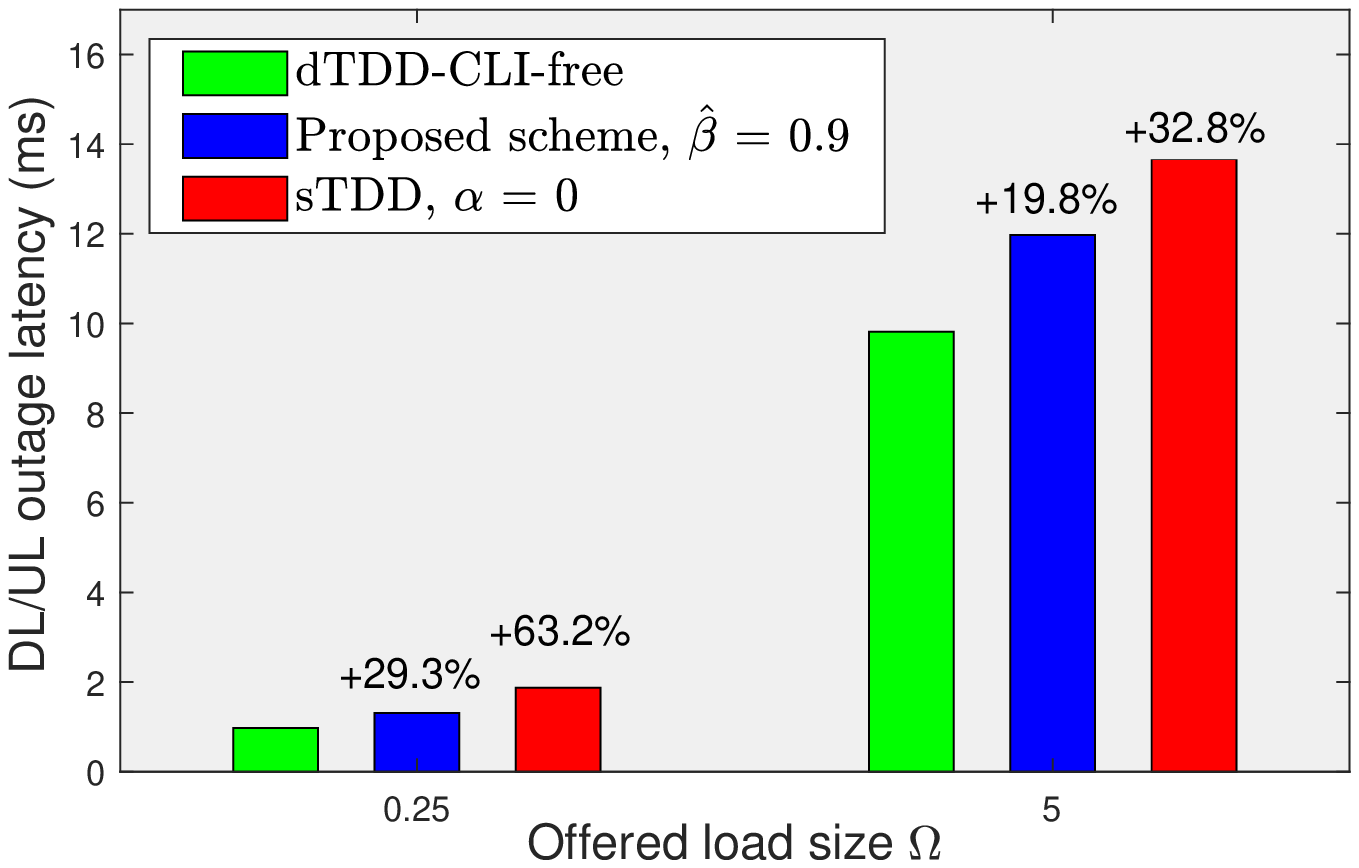}
\par\end{centering}
\centering{}{\small{}Fig. 6. Comparison of the URLLC latency performance
with various offered load levels $\varOmega$. }{\small \par}
\end{figure}

\end{document}